**Reversible and irreversible trapping at room temperature in poly(thiophene) thin-film transistors**

A. Salleo, F. Endicott and R. A. Street

Palo Alto Research Center, 3333 Coyote Hill Road, Palo Alto, CA 94304

## Abstract

We measured the bias stress characteristics of poly(thiophene) semi-crystalline thin-film transistors (TFTs) as a function stress times, gate voltages and duty-cycles. At room temperature, the bias stress has two components: a fast reversible component and a slow irreversible component. We hypothesize that the irreversible component is due to charge trapping in the disordered areas of the semiconductor film. At low duty-cycle (<2%), the fast bias stress component is reversed during the off-part of the cycle therefore the observed $V_T$ shift in only caused by irreversible trapping. Irreversible trapping follows power-law kinetics with a time exponent approximately equal to 0.37. We use these findings to estimate the lifetime of TFTs used as switches in display backplanes.

In addition to high field effect mobility, the practical application of polymer based electronics requires device stability.[1] Polymer thin-film-transistors (TFTs) exhibit degradation under electrical bias (i.e. bias stress), often observed as a threshold voltage ($V_T$) shift after consecutive transfer measurements.[2-7] The effect is caused by positive charge trapped at the dielectric/semiconductor interface while the carrier mobility remains unchanged.[4] However, the charge trapped in accumulation can be quickly released when the gate bias is removed thus a $V_T$ shift is not always observed in the presence of bias stress. Moreover, trapping depends strongly on charge density in the TFT channel therefore $V_T$ shifts may elude experimental observation at lower carrier concentrations.[8,9] As a consequence, bias stress is positively identified only by measuring the TFT current as a function of time during biasing.

We have reported both reversible and irreversible bias stress effects in different polymer TFTs.[10,11] Measurable current decay due to trapping of the mobile charge was observed for gate bias times as short as a few ms. The dependence of the trapping rate on the square of the charge density suggests that the trap states are bipolarons. Understanding the kinetics of charge trapping at short and long timescales is important both to understand its physical origin, and to predict the behavior of TFTs under different biasing cycles and ultimately estimate device lifetime.

Polymer TFTs were fabricated in the bottom gate coplanar configuration using the regio-regular poly(thiophene), PQT-12, as the semiconductor.[12] The substrates were heavily doped Si wafers with a 100 nm layer of thermally grown oxide. After annealing at 410K, the mobility ranged from 0.04 up to 0.1 $cm^2/V.s$. X-ray diffraction measurements indicated the presence of (100)-textured polymer crystallites in the film.[13]

Data consistency and repeatability was verified by measuring the same devices many times. A brief (1 hour) anneal at 410K in vacuum in between measurements reversed accumulated bias stress.

Continuous (DC) and pulsed measurements were performed at room temperature, in vacuum (P<0.2 mTorr) to eliminate effects of the environment. In the DC mode, a constant gate and drain bias was applied while the current was measured. The bias was removed after a given stress time and the current was measured during short (~25 ms) gate pulses applied every ten seconds to measure the recovery rate of the device. In the pulsed mode, the electrical stress was applied with pulses of negative gate bias and the gate voltage was otherwise kept at 0 V. The length of the gate pulses and the driving duty-cycle were varied. The specific choice of off-voltage is not significant because no difference in bias stress reversal was observed when the gate was kept at a positive voltage rather than 0 V. In the linear regime, the $V_T$ shift is related to the change in current by:

$$|\Delta V_T|(t) = \left(1 - \frac{I(t)}{I_0}\right)\left|V_G - V_T^0\right| \quad (1)$$

where $|\Delta V_T|(t)$ is the $V_T$ shift at time t, $I(t)$ is the output current at time t, $I_0$ is the current at time t=0 and $V_T^0$ is $V_T$ at t=0. Experiments confirm that the mobility is unchanged.

Figure 1a shows DC stress-recovery curves for three different stress times. For short stress times (t<100 s) nearly all the trapped charge is quickly recovered when the gate bias is removed. This result is in agreement with our observation that no $V_T$ shift is observed during consecutive measurements of the transfer characteristics of TFTs, which typically take ~30s each.[10] For longer stress times however, only a fraction of the trapped charge is recovered when the gate bias is removed leading to an irreversible $V_T$ shift ($\Delta V_T$). The inset to Fig. 1a shows the fraction of reversibly trapped charge as a function of stress time.

For a given stress time, the trapped charge is therefore partitioned between "fast" reversible and "slow" irreversible traps. Figure 1b shows the measured fast and slow trapped charge density as a function of gate voltage for a stress time of 4000 s. As the gate bias increases, an increasingly large fraction of the total trapped charge is found in slow traps, reaching about 30% at 60V bias.

In order to understand the effect of discontinuous biasing on the electrical stability of TFTs, we measured reversible and irreversible charge trapping as a function of the gate voltage duty cycle. During the on-part of the cycle, the transistors are subject to both reversible and irreversible trapping. During the off-part of the cycle however, charge trapped in "fast" traps is partially detrapped and a fraction of the $\Delta V_T$ is recovered.

As expected, Fig 2a shows that after a fixed bias time, the $|\Delta V_T|$ decreases monotonically with decreasing duty cycle from DC to 2%. The results confirm that at low duty cycle, the longer rest time between stress pulses allows more charge in "fast" traps to be released. For the same reason, at constant duty-cycle, a longer pulse-length leads to less bias stress because of the longer off-time between pulses. The data show that the final $V_T$ associated with irreversible trapping is independent of stress duty cycle and bias pulse length. Therefore, at a given gate voltage, the total amount of charge trapped in "slow" or irreversible traps depends only on the total gate-on time. At the 2% duty cycle shown in Fig. 2a, charge trapped in "fast" traps is completely released during the rest period of the biasing cycle. In these conditions, $V_T$ shifts during operation are purely irreversible and when the biasing cycle is stopped, no $V_T$ recovery is observed.

Operating the transistors at low duty-cycles therefore eliminates entirely the effects of "fast" trapping and allows a selective measurement of the irreversible trapping. Such measurements of the time dependence of the irreversible $V_T$ shift at different gate voltages, are shown in Figure 2b. At higher gate voltages, longer rest times were required (e.g. duty cycle<0.5%). The irreversible $V_T$ shift is approximately fitted with a power law relationship:

$$\left|\Delta V_T\right| = A\left(V_G - V_T^0\right)^n t^\gamma \tag{2}$$

where $\gamma$~0.35-0.45 respectively at high and low $V_G$, and $n$ is about 2. The fitted curves are shown in Fig. 2b and the fit parameters are given in the caption. A similar power law with $\gamma$~0.33 is observed in a-Si:H.[14]

The results of Fig 2a show that irreversible trapping does not depend on the amount of free charge (i.e. device current) but only on the total amount of charge in the channel (i.e. gate bias) and bias time. This observation suggests that during biasing there is a conversion of charge from "fast" traps towards "slow" traps.

We observe that the irreversible trapped charge is entirely released within a few minutes upon illumination with visible light, and similar reversal by illumination was observed earlier in a polyfluorene-co-bithiophene semiconductor (F8T2).[4] The trapping kinetics were also nearly identical for devices processed on PECVD $SiO_2$ and thermal $SiO_2$ gate dielectric.[11] Both observations suggest that charge trapped in "slow" traps is located in the polymer semiconductor and not at the polymer/dielectric interface or inside the dielectric.

Irreversible traps are characterized by slow capture kinetics and their stability implies a large binding energy of the trapped charge. We suggest that the slow and fast stress effects may be related to different local structures of the polymer semiconductor. For example, the high binding energy slow states may be located in highly disordered or amorphous areas of the film where the conjugation of the polymer backbone is poor, while the fast states arise from the ordered regions. The HOMO states of the amorphous regions of the film are pushed to lower energy compared to those of the ordered areas of the film. The energetic barrier for hole injection from the ordered regions into the disordered regions of the film is consistent with the slow trapping kinetics. On the other hand, once the holes have entered the disordered regions, they must successively fall in a deep trap state or they would otherwise eventually hop out and not give rise to a permanent $V_T$ shift. These deep traps could be bipolaron states as well: the strong localization of holes in the disordered regions of the polymer film increases the binding energy of bipolarons compared to the crystalline areas. The exponent n=2 in Eq. 2 is consistent with the bipolaron hypothesis, however more experimental data is needed to confirm it.

During operation of a display backplane, the pixel transistors are kept in their on-state for a time equal at most to the frame time divided by the number of lines, $N$, in the display. Thus, the highest duty cycle of transistors in display backplanes is equal to $1/N$. In these operating conditions, reversible traps play a minor role and bias stress is dominated by "slow" trapping in irreversible traps.

The display lifetime, $T_D=N\tau_L$, can be estimated, based on the TFT lifetime, $\tau_L$, for irreversible trap formation given in Figure 2b, the maximum duty cycle 1/N and a failure criterion that defines the maximum tolerable current drop of the transistor $\left(\frac{I}{I_0}\right)_M$ or equivalently, the maximum shift $\Delta V_T^{\max}$. Combining Eqs. (1) and (2) and approximating $1/\gamma$~3, we obtain:

$$T_D = N\left[\frac{1-\left(\frac{I}{I_0}\right)_M}{A\left(V_G - V_T^0\right)^{n-1}}\right]^{\frac{1}{\gamma}} = N\left[\frac{\left|\Delta V_T^{\max}\right|}{A\left(V_G - V_T^0\right)^n}\right]^{\frac{1}{\gamma}} \propto \left|\Delta V_T^{\max}\right|^3 \left(V_G - V_T^0\right)^{-6}. \quad (3)$$

Figure 3 shows the calculated $T_D$ using Eq. (3) for three values of $\left(\frac{I}{I_0}\right)_M$ : (90%, 50% and 30%), and a gate voltage of -20V. The calculation assumes the display is run continuously; the lifetime will be longer for intermittent operation. Figure 3 highlights the power-law behavior of Eq. (5): $T_D$ is small if only a 10% drop in current is allowed, but increases by two orders of magnitude when $\left(\frac{I}{I_0}\right)_M$ is further decreased from 90% to 50%, when the lifetime is ~3,000 hours for a display with 1000 lines. On the other hand, the lifetime gain is only another factor of two when $\left(\frac{I}{I_0}\right)_M$ is decreased from 50% to 30%. We have not measured the stability of the irreversible traps for this long time and if they reverse slowly, the lifetime would be extended. The lifetime is a strong function of gate voltage so that applications requiring particularly high voltages will have much shorter lifetime.

In conclusion, room-temperature bias stress in PQT-12 TFTs has a "fast" reversible component and a "slow" irreversible component. During pulsed operation, the "slow" component becomes predominant as the duty cycle is reduced. The irreversible "slow" trapping depends only on the integrated gate-on time. We used this property to estimate for the first time the lifetime of transistors used in low duty-cycle operation such as in display pixels. We hypothesize that “slow” trapping occurs in the disordered area of the polymer film. Further studies are needed to confirm this hypothesis and determine the exact nature of the trap states, which may clarify the origin of the trapping kinetics.

**Acknowledgments:** The authors gratefully acknowledge B. S. Ong and the Xerox Research Centre of Canada for providing the PQT-12 material, W. S. Wong, A. C. Arias, M. L. Chabinyc, J.-P. Lu and R. B. Apte of PARC for helpful discussions. This work is partially supported by the Advanced Technology Program of the National Institute of Standards and Technology (contract 70NANB0H3033)

**List of figures**

Figure 1: Normalized current ($V_{DS}$=-1V, $V_G$=-20V) as a function of time for different DC stress times (a). Ratio of released charge to total trapped charge as a function of DC stress time (inset). Partition of trapped charge between reversibly and irreversibly trapped charge as a function of gate voltage after stressing the devices for 4000 s. (b).

Figure 2: $V_T$ shift calculated according to Eq. 1 as a function of time for different pulse-lengths ($V_G$=-20V, $V_{DS}$=-1V) and duty cycles (a). The time units are defined as follows. For t<2000s, the abscissa is the gate on-time, for t>2000s, the abscissa is the real time.

$V_T$ shift measured with low duty cycle biasing as a function of gate-on time for different gate voltages (b). For all the curves shown in the figure, the $V_T$ recovery measured after the biasing cycle was <0.5V. The parameters used to fit the curves to Eq. 2 were A~$8.10^{-4}$, n=2, $\gamma$=0.37 and $\left|V_T^0\right|$ was extracted from transfer measurements prior to stressing the device. $\left|V_T^0\right|$ varied between 4 V and 6 V.

Figure 3: $T_D$ as a function of N according to Eq. (5). The parameters used to calculate $T_D$ were extracted from the fit of the $V_G$=-20V curve of Fig. 2b and are as follows:

A=$7.6.10^{-4}$, n=2, $\gamma$=0.37 and $\left|V_T^0\right|$=6V.

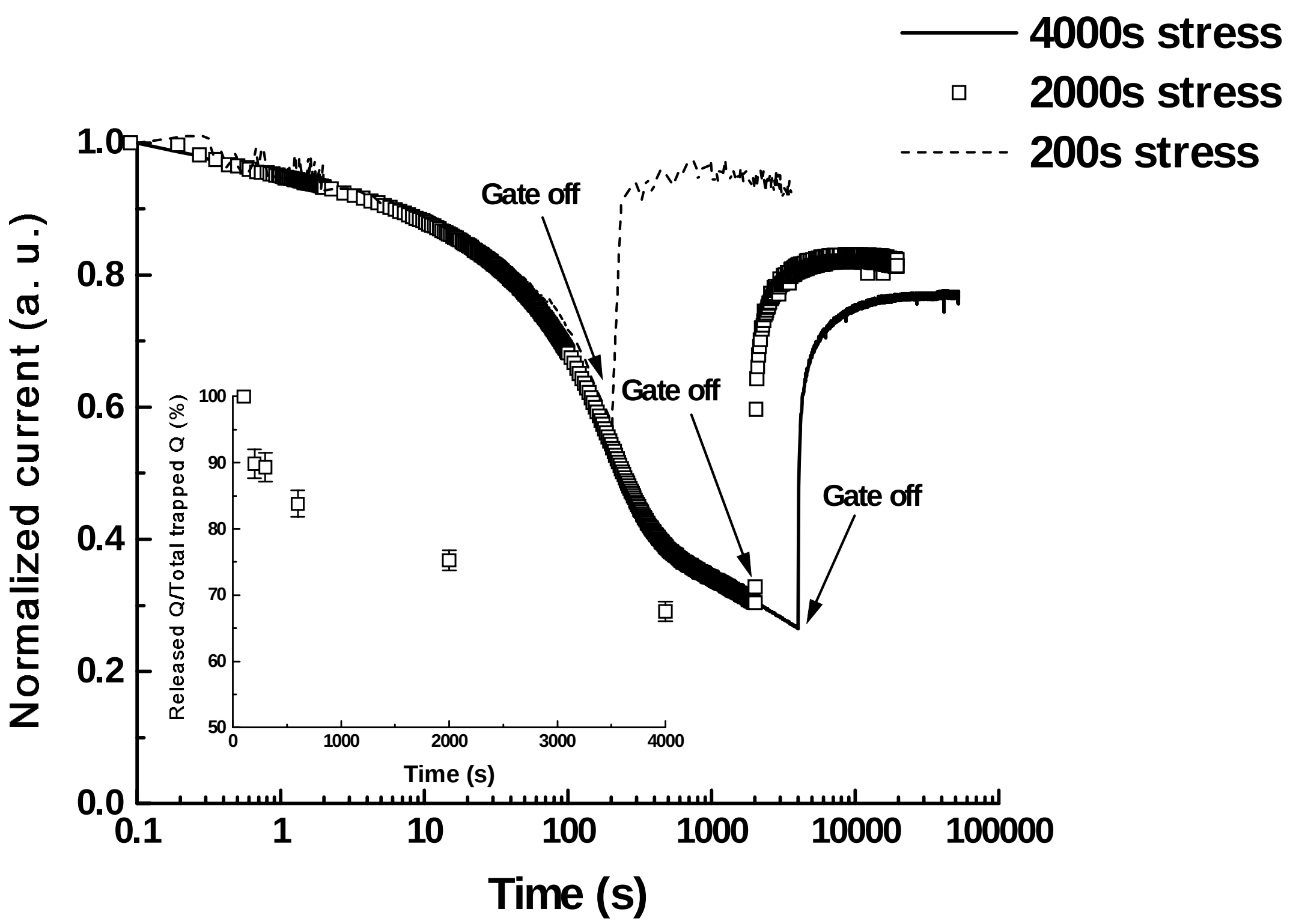

Figure 1a

A. Salleo, F. Endicott and R. A. Street, "Reversible and irreversible trapping at room temperature in poly(thiophene) thin-film transistors".

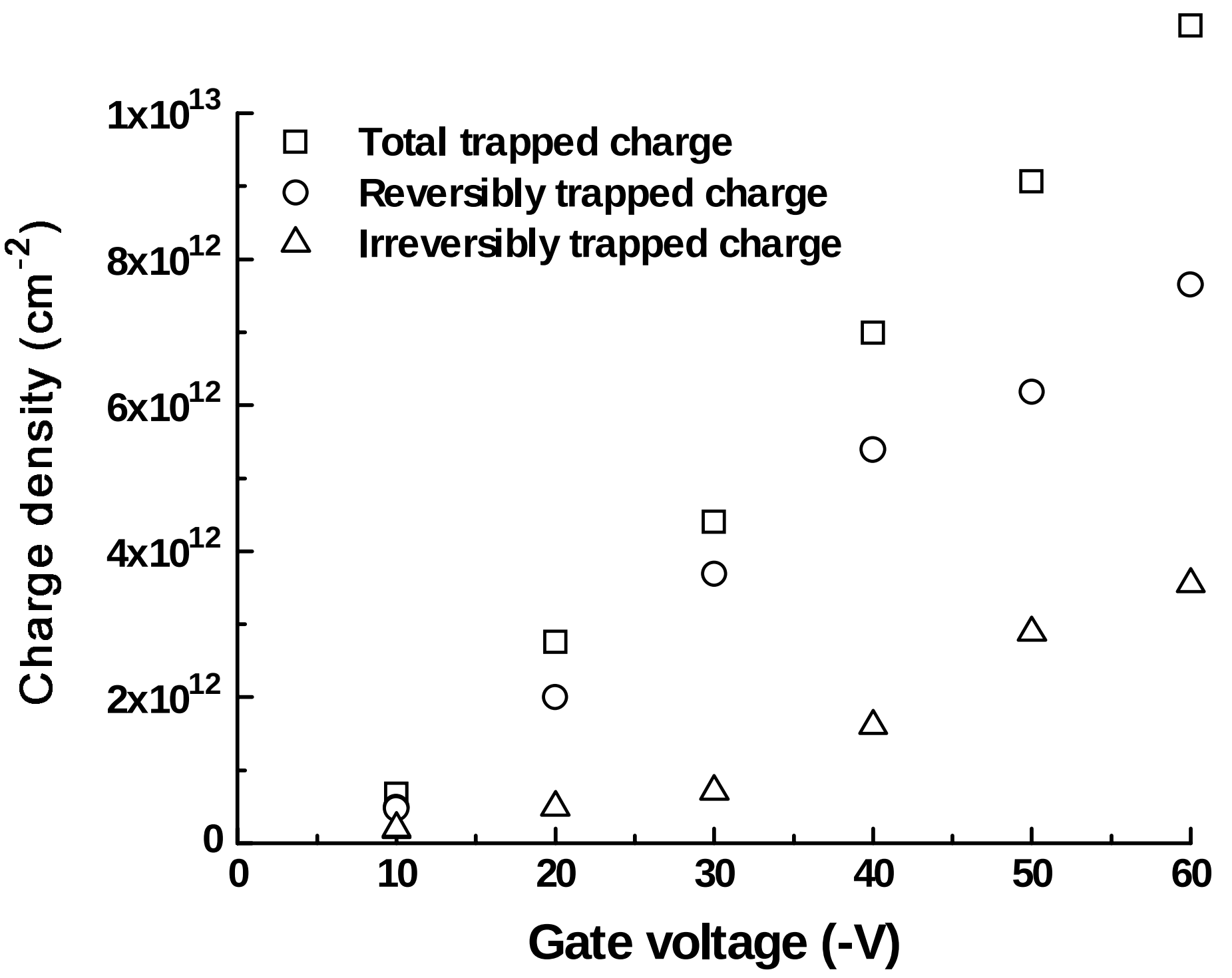

Figure 1b

A. Salleo, F. Endicott and R. A. Street, "Reversible and irreversible trapping at room temperature in poly(thiophene) thin-film transistors".

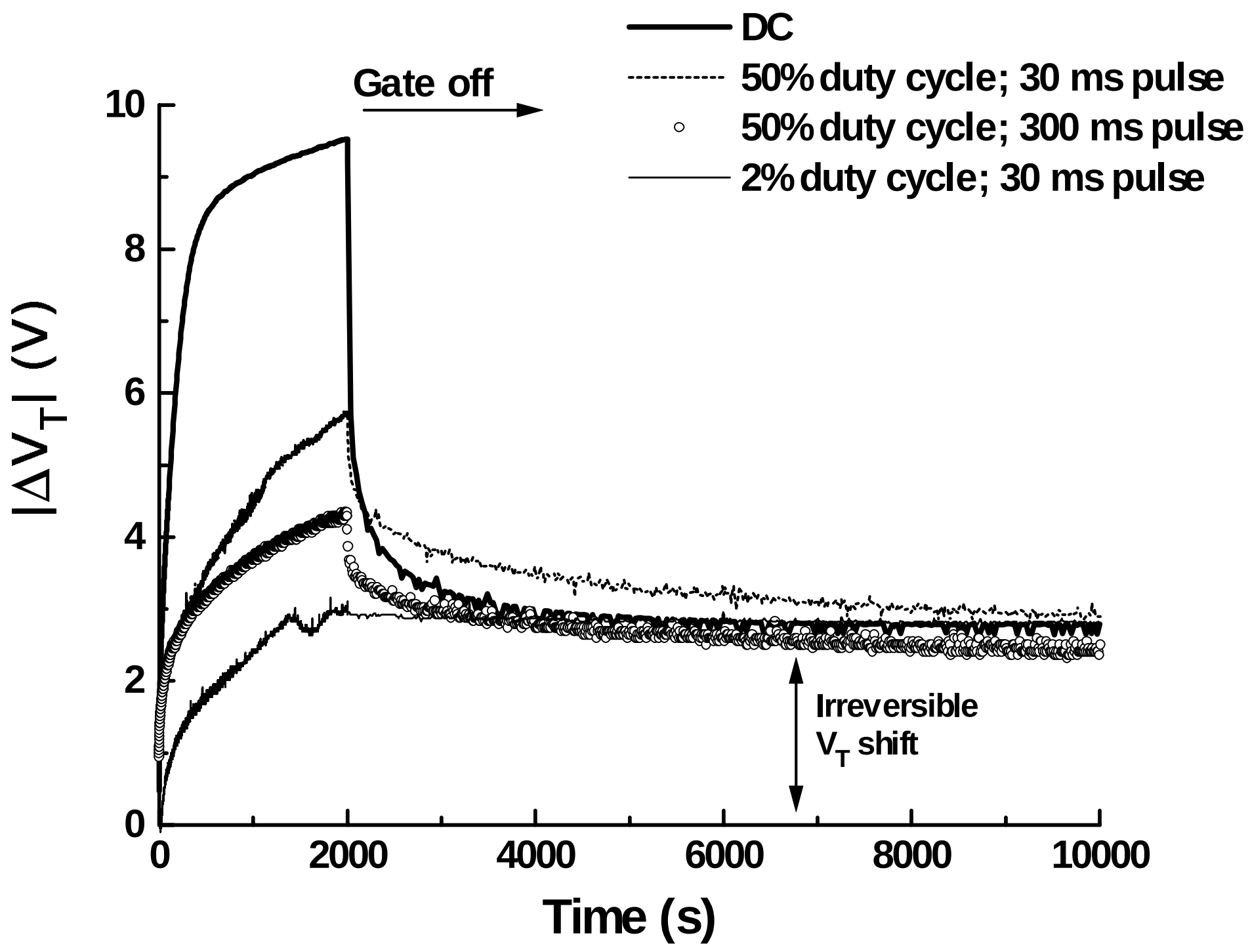

Figure 2a

A. Salleo, F. Endicott and R. A. Street, “Reversible and irreversible trapping at room temperature in poly(thiophene) thin-film transistors”.

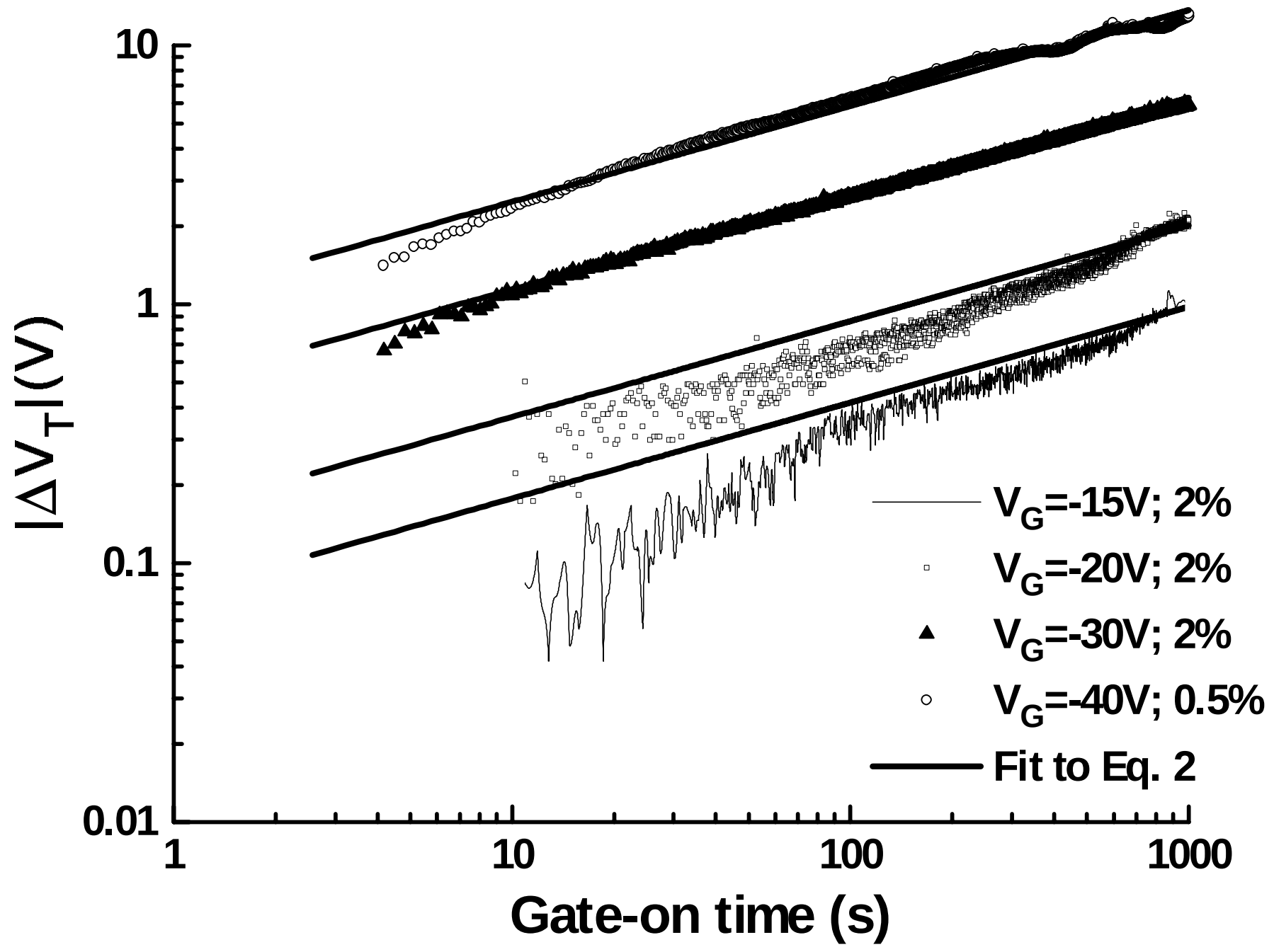

Figure 2b

A. Salleo, F. Endicott and R. A. Street, “Reversible and irreversible trapping at room temperature in poly(thiophene) thin-film transistors”.

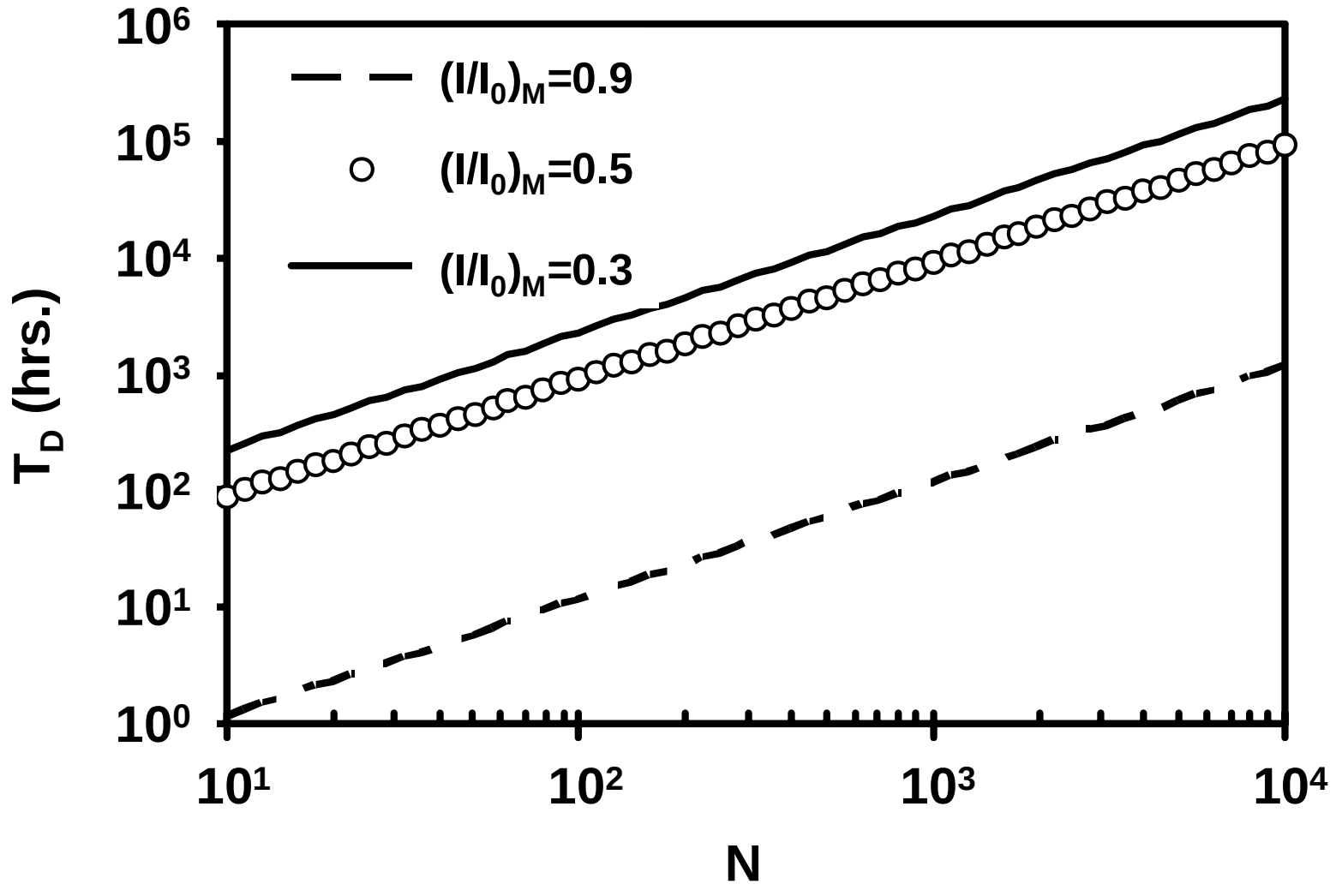

Figure 3

A. Salleo, F. Endicott and R. A. Street, "Reversible and irreversible trapping at room temperature in poly(thiophene) thin-film transistors".